\documentclass[prd,aps,twocolumn,showpacs,superscriptaddress]{revtex4}

\usepackage{graphicx}
\usepackage{amssymb}

\newcommand{\beq}{\begin{equation}}
\newcommand{\eeq}{\end{equation}}
\newcommand{\bea}{\begin{eqnarray}}
\newcommand{\eea}{\end{eqnarray}}

\begin{document}

\title{Deconfinement transition dynamics and early thermalization in QGP}
\author{E. T. Tomboulis}
\email{tombouli@physics.ucla.edu}
\author{A. Velytsky}
\email{vel@physics.ucla.edu}
\affiliation{Department of Physics and Astronomy, UCLA,
Los Angeles, CA 90095-1547}

\date{\today}

\begin{abstract} 
We perform SU(3) Lattice Gauge Theory simulations of the deconfinement 
transition attempting to mimic conditions encountered in heavy ion 
collisions. Specifically, we perform a sudden temperature quench across 
the deconfinement temperature, and follow the response of the system in 
successive simulation sweeps under spatial lattice expansion and temperature 
fall-off. In measurements of the Polyakov loop and structure functions 
a robust strong signal of global instability response is observed through 
the exponential growth of low momentum modes. 
Development of these long range modes isotropizes the system which 
reaches thermalization shortly afterwards, and enters a stage   
of quasi-equilibrium expansion and cooling till its return to 
the confinement phase. The time scale characterizing full growth 
of the long range modes is largely unaffected by the conditions of spatial 
expansion and temperature variation in the system, and is much shorter 
than the scale set by the interval to return to the confinement phase. 
The wide separation of these two scales is such that it naturally results 
in isotropization times well inside 1 fm/c. 
\end{abstract} 
\maketitle

\section{Introduction}

Recent heavy ion collision experiments at RHIC have produced 
a wealth of data on hadron spectra and their anisotropies, in particular 
the magnitudes of radial and elliptical flows.   
This data reveals, perhaps unexpectedly, coherence in 
particle production and strong collective flow phenomena. 
It turns out that about 99\% of the single hadron data ($p_T 
< 1.5$ GeV) are very well described by the hydrodynamics of 
a near-perfect fluid, provided the initial condition of  
very rapid thermalization (in $\sim 0.5 fm/c$) is introduced  
\cite{Heinz:2004pj} - \cite{Kolb:2003dz}.        
This strongly indicates that the quark-gluon plasma (QGP) 
formed at RHIC energies is a strongly coupled fluid.

At asymptotically high temperatures above the deconfinement $T_c$, where 
the running coupling $g(T)$ is small,  
QCD is well-described as a 
gas of weakly coupled quasi-particles, and has been much 
studied by perturbative techniques. 
At the energy densities achieved in the high energy heavy ion collisions, 
however, perturbative treatment of the equilibration process 
appears not to be applicable. Various estimates of 
thermalization times based on perturbative scattering processes   
have been obtained, e.g. in the so called parton-cascade approach 
to the time evolution of hard partons, or the bottom-up scenario 
\cite{MG}, \cite{BjV}, \cite{BMSS} incorporating saturation picture 
(see \cite{Kar}, \cite{IV} for review) initial conditions. 
They all result into 
thermalization times much longer than those needed by 
the hydrodynamical simulations.  It has been argued that this is  
a generic feature of any dynamical  evolution based only   
on perturbative scattering processes \cite{Kov}. 

Even within a weak coupling analysis, however, non-perturbative 
effects may contribute to the dynamics. It has been pointed out that, 
in a plasma with an anisotropic hard parton distribution, instabilities 
may develop in soft gauge field modes 
generated within the linear response approximation 
\cite{ALM} - \cite{H}. It has been argued that these small-amplitude 
unstable modes 
are not stabilized by (non-Abelian) non-linearities; and thus can  
grow to amplitudes large enough to contribute $O(1)$ 
fraction to the total energy density, and drive isotropization 
of the hard modes faster than any hard collision equilibration 
time scale \cite{ALMY}.

Various investigations of the evolution of these semiclassical 
instabilities have been carried out probing beyond the linear regime, 
and, in particular, within the hard-loop effective action \cite{Rebh1} -  
\cite{Rebh2}. They generally indicate that such instabilities indeed persist 
when non-Abelian non-linearities are taken into account within 
the weak coupling regime.  
At some point, however, other non-perturbative dynamics 
at strong effective couplings defined at scales appropriate to  
the nonlinear interaction of such growing long range modes 
must enter. Still, consideration of 
such potential instabilities properly point to a basic underlying 
question which should be addressed from a more  
general point of view.

When systems are driven far from equilibrium by sudden changes in 
external conditions, the approach to a new 
equilibrium state involves, in general, complex nonequilibrium 
processes. Such situations arise, for example, in  
a `quench' to the metastable region across a first-order 
transition boundary, or from a one-phase region to   
the  multiphase coexistence region 
across a second-order transition.  
The study of the dynamics of such far-from-equilibrium 
processes is still at an early stage of development. 
Nonetheless, from many studies, 
mostly in condensed matter physics systems, two broad classes of responses 
have been roughly identified. 

Immediately after a rapid quench, the state of the 
system, as characterized by the appropriate order 
parameter, is nearly identical  
to the state before. The fields must then 
find some way to adjust toward values appropriate to the conditions 
after the quench.  
One way the system may decay towards equilibrium is by the excitation 
of finite amplitude localized fluctuations that may grow or 
coalesce as in nucleation, or interact in other ways over time. This 
typically indicates that the system finds itself in some sort 
of metastable state.   
Another way is by the immediate development of 
a spatially modulated order parameter whose amplitude grows continuously 
from zero throughout the sample. Spinodal decomposition is a prime example 
of this type of response, and the term is often used loosely to 
generically denote such globally unstable behavior. 
It should be pointed out that the boundary between these two 
rough classes is not sharp in all cases.

At RHIC heavy-ion collisions in the central rapidity region achieve 
the sudden deposition of energy densities reaching 
$\sim 30 GeV/fm^3$.  
The first most basic question that must be posed then is: {\it 
which general type of dynamic response is characteristic of 
a rapid transition across the confinement-deconfinement boundary in QCD?} 
In the case of heavy-ion collisions, the 
question must be further qualified by the inclusion of 
the effects of rapid expansion and temperature variation, as well 
as finite volume. This is the issue we explore in this paper.

To get some intuition, we first investigate this question briefly 
within an effective action approach in Section 2. Indeed, 
much of the current understanding of the early time evolution of systems 
out of equilibrium has been obtained by investigating classes of 
stochastic equations that are natural dynamical (time dependent) 
generalizations of the Landau-Ginzburg (LG) effective action models 
of the static (equilibrium) theory (\cite{gunton:1985}, \cite{Chaikin:1997}). 
In the case of the 
confinement - deconfinement transition, the 
relevant order parameter is the Polyakov loop (Wilson line), and 
LG effective actions for it have been considered in 
\cite{Pisarski:2000eq} - \cite{HKW}. Corresponding 
dynamical model generalizations  can then be used to 
examine the question posed above. We consider the predictions of such a 
model briefly in Section 2 below.

Though the effective model approach often proves valuable, what is 
ultimately needed is an ab initio treatment in the  
full nonperturbative formulation of the exact theory, i.e.  
lattice gauge theory.  Unfortunately, there is no established 
simulation formalism for directly extracting physical 
properties in non-equilibrium real-time evolution in quantum field theory. 
What one can do, however, is 
mimic Minkowski real-time dynamics by Glauber stochastic dynamics evolution. 
Thus starting with the system in thermalized equilibrium, one 
performs a temperature quench and follows the system, 
over successive simulation sweeps,   
on its path toward regaining equilibrium.  
Though this cannot be directly identified with the exact real-time   
evolution, it is known from many studies, mostly  of condensed matter 
systems, to accurately reflect it. At the very least, 
the method provides a consistent picture of the basic features of the 
system's real-time response. It has been extensively and successfully used 
for many systems  exhibiting, in particular, first-order transitions.  
Indeed, in studies of binary alloys and 
binary fluids it has been found to reproduce the experimentally 
observed behavior in quantitative detail \cite{gunton:1983}.  
In the case of gauge theories, the method was first 
used in the pioneering studies in \cite{Miller:2000mr,Miller:2000pd,
Miller:2001ym}. 
More, recently, such studies of the dynamics of phase transitions in 
spin and gauge theory systems were undertaken and much extended 
in \cite{Bazavov:2004wc,Berg:2003mn,Berg:2003hc,Berg:2004qb, 
Velytsky:2002fn}.

In this paper, building on these previous gauge theory studies, we 
consider $SU(3)$ gauge theory under conditions  
mimicking  the situation encountered in heavy-ion collisions. 
This we do in Section 3 which constitutes the main part of 
the paper. Specifically, we examine the response after 
a sudden quench into the deconfinement phase under varying 
conditions of spatial expansion and 
temperature variation. In this first investigation, we consider 
only pure $SU(3)$ gauge theory, i.e. no quarks, as the deconfinement 
transition is driven by gluonic dynamics. 
In simulation measurements of structure functions and Polyakov loops, 
we follow the evolution of the system from the quench till its 
return to the confinement phase. The main outcome of our study 
is that two relevant, widely separated  
time scales emerge. First, a strong and robust signal of rapid growth of very 
long range modes is observed after the quench.  
Most importantly, the time scale of full development of these 
low momentum modes is essentially unaffected  by the presence, over a 
wide range of parameters, of spatial expansion and temperature 
variation in the system. 
The development of these modes `isotropizes' the system, which 
reaches full quasi-equilibrium shortly afterwards as signaled 
by the full decay of the structure function.  It thus enters 
a stage of expansion and cooling, characterized by a second, much 
longer time scale, till its return to the confinement phase. 
The wide separation of these two scales accords well with 
what is seen in heavy-ion collision experiments.  
Having arrived at this robust qualitative picture, we also 
attempt to make a more quantitative comparison of what is seen in 
these simulations to hydrodynamical phenomenology (subsection 3.3). 
There are certainly uncertainties here, not least of which is 
the fact that we do not include fermions which can make a significant 
contribution especially at the late stage near the return to confinement.  
Still, one finds that the separation of scales is such that, for any 
reasonable choice of parameters, 
isotropization times well inside $1 \ fm/c$ result naturally.

Finally, in Section 4  we briefly discuss our conclusions and 
directions for further work.

\section{Effective action models}

Effective action models allow one to build simple theories of initial
time evolution of systems driven out of equilibrium by a quench. 
Such theories can be build for general classes of models. A simple
linear theory (Cahn-Hilliard) was first proposed for models of type B with
conserved order parameter \cite{Cahn:1958}. The generalization to
models of type A with non-conserved order parameters is straightforward,
see, for example, \cite{Berg:2004qb,Velytsky:2004}. For a general
overview see e.g. \cite{gunton:1985,Chaikin:1997}.

Though such effective models are not our primary focus in this paper, it is 
useful to briefly consider them as they provide useful insight into 
the possible behavior of the system that can be checked against the 
outcome of simulations in the actual gauge theory. 
For $SU(3)$ gauge theory the low energy degrees of freedom are represented
by Polyakov loops, and standard potential models for them 
are known and well studied \cite{Pisarski:2000eq,W,HKW}.   
Adopting the potential for the Polyakov loop $l$ (a complex quantity) in 
\cite{Pisarski:2000eq}
\begin{equation}
{\mathcal{V}}(l)=\left(-\frac{b_{2}}{2}|l|^{2}-
\frac{b_{3}}{6}(l^{3}+(l^{*})^{3})+
\frac{1}{4}(|l|^{2})^{2}\right)b_{4}T^{4} \;, \label{Polyak-pot} 
\end{equation}
the coupled set of Langevin equations is
\begin{equation}
\frac{\partial l}{\partial t}=-\Gamma\frac{\delta S}{\delta l^{*}}+\eta, 
\quad \mbox{and its}\quad c.c. 
\label{eq:langevin1}
\end{equation}
Here $S$ is the standard Landau-Ginzburg action 
\beq
S=\int d^{3}x\left(\frac{1}{2}|\partial_{i}l|^{2}+{\mathcal{{V}}}(l)\right)
\;,\label{LGact}
\eeq
$\Gamma$ is the response coefficient, which defines the relaxation time scale 
of the system,
and $\eta$ is a noise term. We will ignore the noise term, since 
it can be shown that it does not affect the resulting rate of growth of 
fluctuations \cite{F1}.

We are interested in the  fluctuations of the Polyakov loop $l$ around
some average value $l_{0}$: 
\beq
l(\vec{{r}},t)=l_{0}+u(\vec{{r}},t),  
\eeq
Using  (\ref{Polyak-pot}) and (\ref{eq:langevin1}), the system of equations 
for the fluctuations is
\beq
\frac{\partial u}{\partial t}=-\Gamma\left[-\frac{1}{2}\nabla^{2}u+
\left(c_{1}u+c_{2}u^{*}+c_{3}\right)\right] \quad 
\textnormal{and its} \quad c.c. \;, \label{flucteqs}
\eeq
where $c_{1}=(-b_{2}/2+|l_{0}|^{2})b_{4}T^{4}$, 
$c_{2}=(1/2l_{0}^{2}-b_{3}l_{0}^{*})b_{4}T^{4}$
and $c_{3}=(-b_{2}/2l_{0}-b_{3}/2l_{0}^{*2}+1/2l_{0}^{2}l_{0}^{*})b_{4}T^{4}$
are complex numbers. Note that $c_{1}=c_{1}^{*}$.

We solve for the Fourier transforms $u(\vec{{k}},t)$ and $v(\vec{{k}},t)$
of the fluctuations $u(\vec{{r}},t)$ and $u^{*}(r,t)$, respectively, 
which, from (\ref{flucteqs}) satisfy: 
\begin{eqnarray}
\frac{\partial u(\vec{{k}},t)}{\partial t}+
\Gamma\left[(\frac{1}{2}k^{2}+c_{1})u(k)+c_{2}v(k)\right] & = & c(k)
\nonumber \\
\frac{\partial v(\vec{{k}},t)}{\partial t}+
\Gamma\left[(\frac{1}{2}k^{2}+c_{1}^{*})v(k)+c_{2}^{*}u(k)\right] 
& = & c^{*}(k) \;,  
\end{eqnarray}
where $c(k)=c(-k)=-\Gamma c_{3}\sum_{\vec{{r}}}
\exp(i\vec{{k}}\cdot\vec{{r}})$. Note also that 
$u^{*}(-\vec{{k}},t)=v(\vec{{k}},t)$.  
The non-zero modes are governed by the homogeneous part of these 
equations. Its eigenvalues
are
\beq
\omega_{1,2}(k)=-\Gamma(\frac{1}{2}k^{2}+c_{1}\pm|c_{2}|),
\eeq
and the eigenvectors are
\beq
S_{1,2}=\left(\begin{array}{c}
\pm c_{2}/|c_{2}|\\
1\end{array}\right).
\eeq
The solution is
\beq
\left(\begin{array}{c} u(\vec{k},t) \\
v(\vec{k},t)
\end{array}\right)
=C_{1}S_{1}e^{\omega_{1}(k)t}+C_{2}S_{2}e^{\omega_{2}(k)t}.
\eeq
Here, an analysis of modes similar to that of the real case (spin systems, 
$SU(2)$) (\cite{Bazavov:2004wc} - \cite{Berg:2004qb}) can be applied.
If $c_{1}\pm|c_{2}|<0$, one will observe exponential growth. 
Spinodal decomposition-like behavior then results if we have 
exponential growth in at least one exponent, i.e. when 
$c_{1}<|c_{2}|.$ The structure function (connected Polyakov 2-point 
function) 
\beq
S(\vec{k},t) = \Bigg< u(\vec{k},t)\,v(-\vec{k},t)\Bigg> 
\eeq
follows similar behavior (Cf. \cite{Bazavov:2004wc}). 

Next, we want to study the effect of spatial expansion. 
We work in new coordinates
of rapidity $\eta=1/2\ln(t+z)/(t-z)$ and proper time $\tau$ : 
\beq
t  =  \tau cosh(\eta)\;, \qquad 
z =  \tau sinh(\eta).
\eeq
Here we assume the $z$-axes to be the axes of expansion (collision).
The change in the corresponding part of the Minkowski metric is 
$dt^{2}-dz^{2}=d\tau^{2}-\tau^{2}d\eta^{2}$, with 
the transverse coordinates left unchanged. Keeping the rapidity constant
results in a constant rate of expansion in the system, the speed of
expansion being  proportional to the distance from the collision center: 
$v=z/t$, with $t$ the time after collision. 
In the new coordinates the metric is similar to the Robertson-Walker 
metric and is defined as $ds^{2}=g_{\mu\nu}dx_{\mu}dx_{\nu}
=d\tau^{2}-a^{2}(\tau)d(\tau_{0}\eta)^{2}-dx_{\perp}^{2}$, i.e. 
\beq
g_{\mu\nu}= \mbox{diag}\,\left(1,\;-1,\;-1,\;-a^{2}(\tau)\right)
\eeq
where $a(\tau)=\tau/\tau_{0}$ is the `scale factor', 
and $\tau_{0}$ is the parameter controlling the rate of expansion.

The obvious naive generalization of the model above 
is to substitute
\[
\nabla^{2}\rightarrow g_{ij}\partial_{i}\partial_{j}\]
in the action (\ref{LGact}), and to consider dynamics in the proper frame.

Eq. (\ref{eq:langevin1}) now gives 
\beq 
\frac{\partial l}{\partial\tau}=-\Gamma\left[-\nabla_{\bot}^{2}l-a^{2}(\tau)\nabla_{\Vert}^{2}l+\frac{\partial V(l)}{\partial l}\right],
\quad \mbox{and its} \quad c.c. \;, 
\eeq
where $\nabla_{\bot}^{2}=\partial_{x}^{2}+\partial_{y}^{2}$ and
$\nabla_{\Vert}^{2}=\partial_{\eta}^{2}$. 
Going through the previous development amounts to the naive 
substitution
$\vec{k}^{2}\rightarrow\vec{k}_{\perp}^{2}+a^{2}(\tau)k_{\Vert}^{2}$
in the eigenvalues: 
\beq 
\omega_{1,2}(k)=-\Gamma\left[\frac{1}{2}(\vec{k}_{\perp}^{2}+a^{2}(\tau)k_{\Vert}^{2})+c_{1}\pm|c_{2}|\right].
\eeq
This now gives a growth of fluctuations governed by a 3rd order polynomial
in $\tau$ in the exponent: 
\begin{widetext}
\beq
\left(\begin{array}{c} u(\vec{k},t) \\
v(\vec{k},t)
\end{array}\right)=C_{1}S_{1}\exp\int_{0}^{\tau}\omega_{1}(\vec{k},\tau^{\prime})d\tau^{\prime}+C_{2}S_{2}\exp\int_{0}^{\tau}\omega_{2}(\vec{k},\tau^{\prime})d\tau^{\prime}.
\eeq
\end{widetext}
Thus we observe that if
\[
\frac{1}{2}(\vec{k}_{\perp}^{2}+a^{2}(\tau)k_{\Vert}^{2})<c_{1}-|c_{2}| \;, 
\]
there is an explosive growth, much faster than in 
the non-expanding case. Let
us for simplicity consider only longitudinal modes. We see that
the critical mode \cite{Chaikin:1997} is 
\begin{equation}
k_{\parallel}^{c}=\frac{2}{a^{2}(\tau)}(c_{1}-|c_{2}|).
\end{equation}
This is rather remarkable dynamics where the critical mode (and all modes 
at momentum scales below it) is moving
in time. During initial time there are more scales involved 
but as time progresses only very large scale regions participate.  
The prediction of growth under expansion by a
higher than linear power exponent is tested against the 
simulation results in the following section.

\section{Simulation study of deconfinement dynamics in $SU(3)$ LGT}

In this section we present a numerical study of the effects
of spatial expansion and varying temperature on the dynamical evolution 
following a sudden quench into the deconfined phase of 
the $SU(3)$ gauge theory on the lattice. We thus try 
to mimic conditions encountered in heavy-ion collisions.  
We use periodic boundary conditions, as we do 
not study the effects of finite size per se. A study of the role of 
the finiteness of the system would certainly be of interest, but 
is left for a future study. Also, in the customary  
heavy-ion collision picture the initial expansion is 
one-dimensional, becoming three-dimensional at later stages of the evolution 
\cite{Bjorken:1983}. Here, as we discuss further below, 
we simplify matters by considering a uniform expansion
in all spatial directions. This is related to isotropic expansion in the 
proper frame at a fixed rapidity value. 
We first study the effect  of such 
expansion; then we add the effect of varying temperature. 

We use a field based heat bath update of the $SU(3)$ fields. 
To accelerate the dynamics there are 2 attempts to update the field per sweep.
This is somewhat different from standard link-based updates; however, it is 
still in Glauber universality class. The time flow is proportional to the 
link-based heatbath with links visited in random order.
Therefore, the peak values of the non expanding quench are slightly different
from previous studies \cite{Bazavov:2004wc} where a heat bath update is 
performed on links visited in systematic order.

We use a space-time anisotropic lattice in order to be able to 
independently control temperature and expansion. 
The anisotropic action is \cite{Karsch:1982ve} 
\begin{equation}
S=-\beta_{\xi}/3\sum_{x}\textrm{Re}\left[\xi^{-1}\sum_{i>j}\textrm{Tr}U_{x,ij}+
\xi\sum_{i}\textrm{Tr}U_{x,0i}\right], \label{action} 
\end{equation}
where $i$, $j$ runs over space-like directions, $\xi=a/a_\tau$ is 
the space-time anisotropy,
$\beta_{\xi}=6/g_{\xi}^{2}$, and $g_{\xi}^{2}=g_{\sigma}\cdot g_{\tau}$ is the
anisotropic coupling. 
Then, by varying two parameters it is possible to carry out expansion 
of the system while maintaining constant temperature, or also 
let the temperature drop thus
allowing for cooling of the expanding plasma. 

\subsection{Spatial expansion} 

Hubble-like uniform expansion of the metric amounts to varying the 
space-like lattice spacing $a$ as  
\begin{equation}
a=a_{0}(1+\frac{\tau}{\tau_{0}}),
\end{equation}
where $\tau$ is the proper time variable, and $\tau_{0}$ is the parameter 
which controls the rate of expansion.

We focus here solely on the role of the expansion. So, after the initial 
quench, we keep the temperature 
$T=1/(N_{\tau}a_{\tau})$ constant throughout the 
expansion by fixing the time-like spacing $a_{\tau}$.
This implies that the anisotropy $\xi=a/a_{\tau}$ traces the
changes in $a$. Thus, assuming zero time anisotropy to be $\xi(\tau=0)=1$, 
one has
\begin{equation}
\xi(\tau)=1+\frac{\tau}{\tau_{0}}.
\end{equation}

Next consider the dependence of the space-like lattice spacing
on the coupling and anisotropy. The one-loop order  renormalization
group relationship is
\begin{equation}
a\Lambda(\xi)=\exp\{-1/(2b_{0}g_{\xi}^{2})\},
\end{equation}
where $b_{0}=11\cdot3/(48\pi^{2})$, and $\Lambda(\xi)$ is dependent
on $\xi$ \cite{Karsch:1982ve} through 
\begin{equation}
\Lambda(\xi)/\Lambda_{E}=\exp\{-(c_{\sigma}(\xi)+c_{\tau}(\xi))/4b_{0}, 
\end{equation}
where $c_{\sigma}(\xi)$ and $c_{\tau}(\xi)$ are known functions. 
Inclusion of the next order terms adds minor corrections for
the range of the couplings and anisotropies used and is 
straightforward. It does, however, complicate 
numerical treatment, since it requires a numerical solving of the 
corresponding equation.

Following the procedure in \cite{Karsch:1982ve} we compute  
$\Lambda(\xi)$ for a range of anisotropy values as listed 
in table \ref{tab:Lambda}.
\begin{table}
\caption{The lattice scale parameter $\Lambda$ dependence on the anisotropy.\label{tab:Lambda}}
\begin{ruledtabular}
\begin{tabular}{|c|c|c|c|c|c|c|c|}
$\xi$& 1& 2& 3& 4& 5& 6& 7\tabularnewline
\hline
$\Lambda(\xi)/\Lambda_{E}$& 1.000& 0.837& 0.801& 0.798& 0.804& 0.812& 0.820\tabularnewline
\hline
$\xi$& 8& 9& 10& 11& 12& 13& 14\tabularnewline
\hline
$\Lambda(\xi)/\Lambda_{E}$& 0.827& 0.833& 0.838& 0.843& 0.847& 0.851& 0.854\tabularnewline
\hline
$\xi$& 15& 16& 17& 18& 19& 20& 21\tabularnewline
\hline
$\Lambda(\xi)/\Lambda_{E}$& 0.858& 0.860& 0.863& 0.865& 0.867& 0.869& 0.871\tabularnewline
\end{tabular}
\end{ruledtabular}
\end{table}
This allows us to estimate the necessary time evolution of $\beta_{\xi}$
\begin{equation}
\beta_{\xi}(\tau)=\beta(0)-6\cdot2b_{0}\log\left[(1+\frac{\tau}{\tau_{0}})
\frac{\Lambda(\xi)}{\Lambda_{E}}\right].\label{eq:betaxi}
\end{equation}
It is important to indicate here that a non-perturbative further 
correction \cite{Boyd:1996bx} need to be applied, when appropriate 
(see subsection 3.3 below). 

\begin{figure}[ht]
\includegraphics[width=\columnwidth]{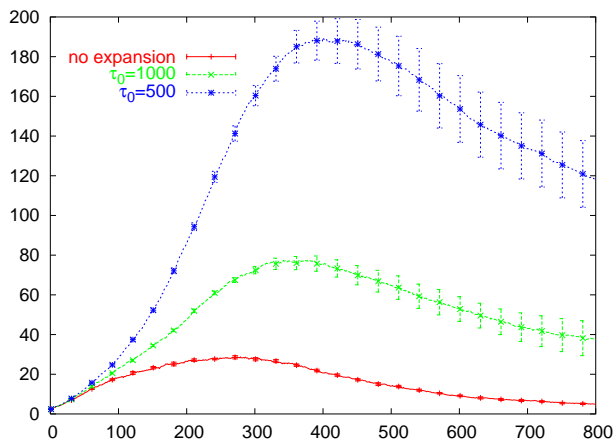}
{\caption{First mode of the structure function at different 
inverse expansion rates 
$\tau_0=500,1000,\infty$. Quench $\beta_{\xi}=5.5\rightarrow5.92$ 
on $16^{3}$x$4$ lattice.
\label{fig:sf1}} }
\end{figure}

Note that in the standard application of anisotropic lattices (such as in
\cite{Karsch:1982ve}) the space-like lattice spacing is not varied; 
the anisotropy is varied by decreasing the time-like lattice spacing. 
This procedure keeps the coupling within the scaling window provided 
the initial coupling is close to the continuum limit.
Here, on the contrary, we keep the time-like spacing
constant (or, later, slowly increase it) as we vary the space-like coupling.
This induces changes in the coupling that may drive its value out
of the scaling regime ($\beta\sim5$). Therefore, our expansion has
to be truncated whenever the value of $\beta$ falls below this cut-off
value. This condition implies that, in order to follow the system evolution 
for longer time, one needs lattices with larger $N_{\tau}$, thus rendering 
the problem more computationally intensive. 

We start simulations on smaller lattices where it is easier to 
gather satisfactory
statistics for highly fluctuating quantities, such as the structure function.
The quench is performed from $\beta_{\xi}=5.5$ to $\beta_{\xi}=5.92$  
on $16^{3}$x$4$ lattice. The latter corresponds to a temperature after 
the quench  $T_{\rm final}=1.57T_c$.  
The phase transition on this lattice at $\xi=1$
is at $\beta=5.6902(2)$. We use jack knife average over 10 bins, each
of 50 configurations. The system is allowed to equilibrate for 200
lattice sweep, and then, after performing a quench, we allow the system
to evolve for 800 sweeps, while measuring several lower modes of the
structure function. We present here only averages of on-axis modes,
such as permutations of $(n,0,0)$ - this is the n-th mode in our
notation. The first modes are presented in Fig. \ref{fig:sf1}. We see that
expansion significantly enhances the response. The faster the expansion
rate, the higher are the peaks. On the other hand the shift in 
the location of the peaks is not as pronounced, an important point 
to which we return below. 

\begin{figure}[ht]
\includegraphics[width=\columnwidth]{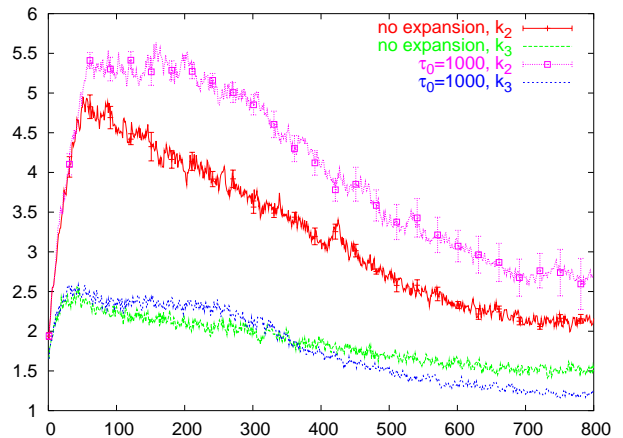}
\caption{Higher ($k_{2}$ and $k_{3}$) 
structure function modes for non-expanding
and $\tau_{0}=1000$ expanding system.\label{fig:sf23}} 
\end{figure}

Next, we look at higher modes of the structure function in the cases 
of no expansion, and expansion at $\tau_{0}=1000$ -- see Fig. \ref{fig:sf23}. 
We see that the second mode shows behavior similar to that observed for the
first mode. The difference, however, is not that significant. The third mode
in the expanding system outgrows the corresponding mode in the 
non-expanding case at early times, but then decreases faster. 
This is an indication of the shift
of the critical mode with time, observed in the linear effective model 
(section 2). At later times the transition proceeds only through 
the lower modes. The error bars for some of the data points are also 
presented in Fig. \ref{fig:sf1} and \ref{fig:sf23}.

\begin{figure}[ht]
\includegraphics[width=\columnwidth]{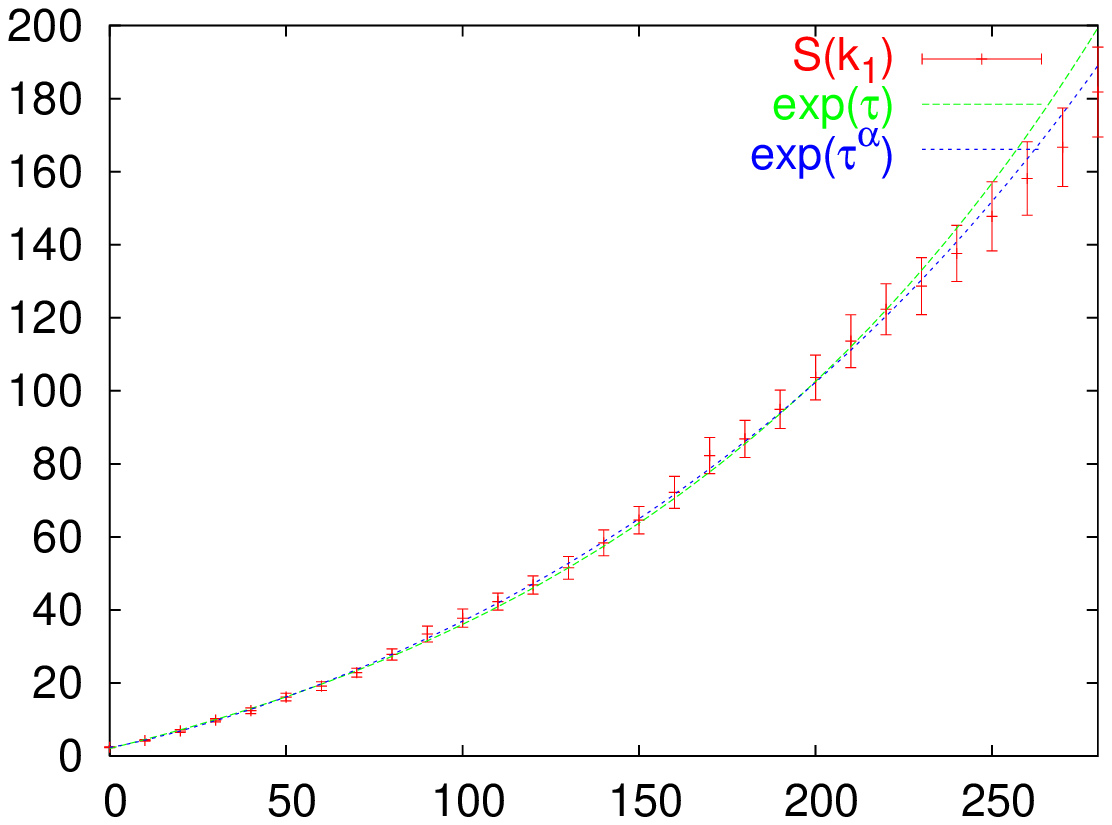}
\includegraphics[width=\columnwidth]{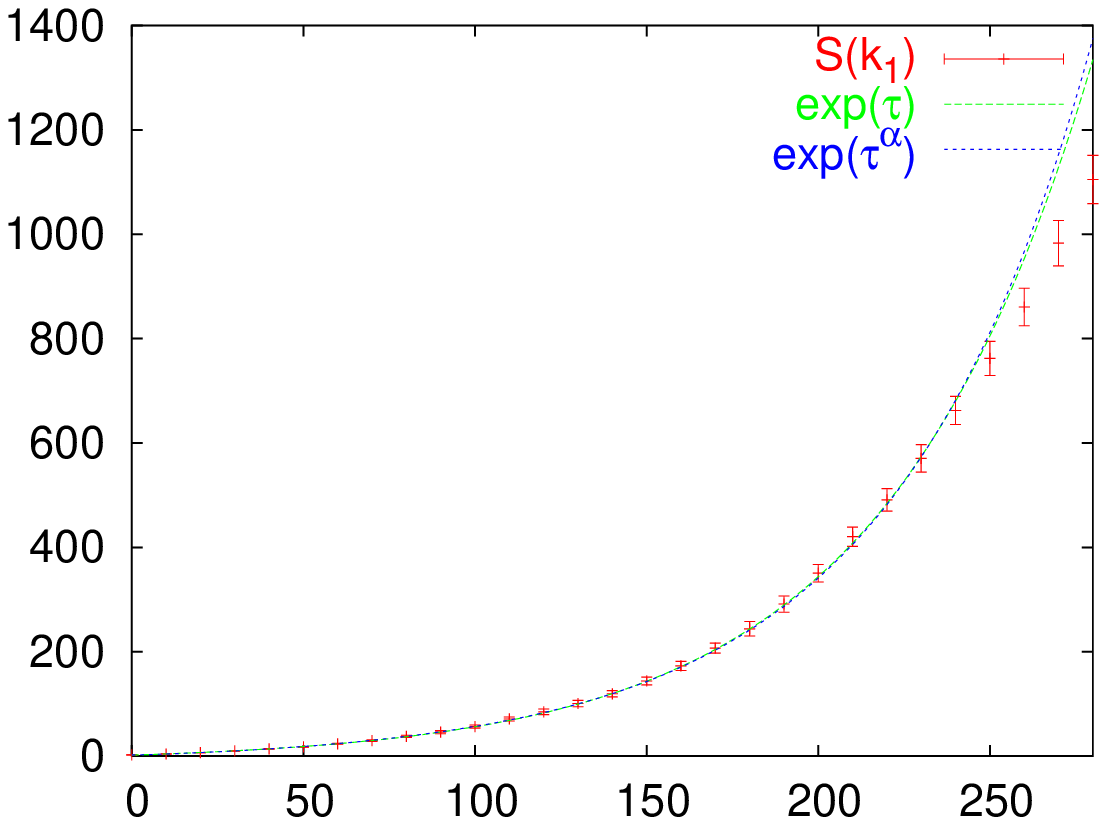}
\caption{Fits to the exponent. Quenches on 
the $32^3$x$4$ lattice  
in non-expanding (top) and $\tau_0=500$ expanding (bottom) systems.
\label{fig:fit}} 
\end{figure}

To make a comparison to the linear effective theory of Section 2, 
we make fits to the exponent for the structure function data 
(Fig. \ref{fig:fit}). We use a $32^{3}$x$4$ lattice
since we know from previous studies \cite{Berg:2004qb,Bazavov:2004wc} 
that the linear response behavior (pertinent to early times) manifests 
itself better on the larger lattices. Contrary to our
expectations from the linear theory of section 2, however, we 
find there no substantial change in the 
behavior of the exponent between the expanding and non-expanding systems.   
A fit to 
\[
	S(\tau)\sim \exp(C\cdot\tau^\alpha)
\]
gives $\alpha=1.22$ for the expanding system, whereas it gives 
$\alpha=1.18$ for the non-expanding system. In Fig. \ref{fig:fit} 
we also show fits to  
$S(\tau)\sim \exp(C\cdot\tau)$, which in fact provides the best fit per 
parameter degree of freedom. Both fits are over the $\tau$ range 
from $0$ to $250$. In any case, there is no substantial deviation 
from exponential growth with linear $\tau$ dependence in the exponent.   
Furthermore, from the figures we observe divergence from  
exponential behavior at later times $\tau>200$. 
Also notable is the enhancement of the signal (by a factor $\sim$ 7) 
in the expanding system as compared to the non-expanding system.  
All this provides a manifestation 
of the limitations of the linear response effective models in Section 2. 
We note that there are 10 times as many points for the fitting 
as on the plot.

\subsection{Expansion accompanied by temperature fall-off} 
In this second part of the numerical study we try to mimic conditions 
similar to those in heavy ion collisions at RHIC. We want to follow 
the evolution of the expanding and cooling QGP.  We let the temperature 
drop as
\begin{equation}
T=\frac{T_{0}}{(1+\tau/\tau^\prime_{0})^{\alpha}}.
\end{equation}
In the fire-tunnel (in the proper frame) the temperature is 
expected to decrease with the proper time $\tau$ as 
$T\sim\tau^{-1/3}$ or slower \cite{Bjorken:1983}. We adopt this 
value of $\alpha$ in the following.

\begin{figure}[ht]
\includegraphics[width=\columnwidth]{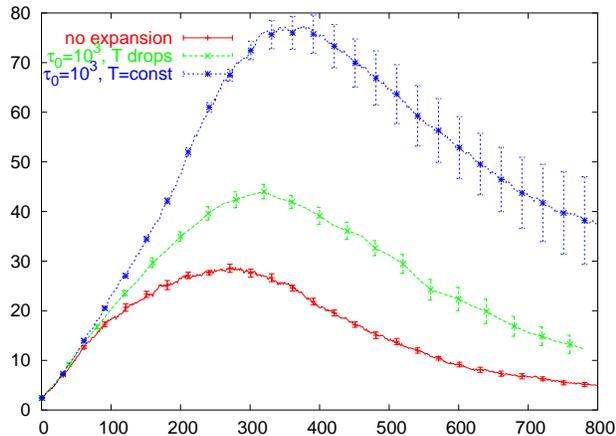}
\caption{Comparison of first mode of structure 
functions including drop in temperature on a $16^3\times4$ lattice 
(same quench as in Fig. 1).
\label{fig:tdrop}}
\end{figure}

We first work on smaller lattices for elucidating the 
main physical features. Therefore, we do not 
set the freeze-out condition (see below) and simply take  
$\tau^\prime_0=\tau_0$.
With the $\alpha=1/3$ choice of temperature evolution, the anisotropy 
evolves as
\begin{equation}
\xi(\tau)=\left(1+\frac{\tau}{\tau_{0}}\right)^{2/3}
\end{equation}
and $\beta$ evolves as in (\ref{eq:betaxi}). We illustrate the effect of 
temperature drop on the structure function 
in Fig. \ref{fig:tdrop}. 
There are two basic conclusions suggested by these 
plots. 

The first is that the temperature evolution drives the system 
back towards the confined phase, but the expansion tends to prevent it   
as evidenced by the different peak heights. These are then two competing 
effects that tend to cancel, so that the return to structure function 
equilibration  occurs at about the same time (here after 
about $700$ sweeps) as for the system in the absence of expansion 
and temperature fall-off. For the slower expansion rates ($\tau_0 > 1000$) and 
faster temperature fall-offs this cancellation effect is even 
more pronounced. 

The second conclusion is suggested by the fact, also present and 
remarked upon  in Fig. \ref{fig:sf1}, 
that the location of the peaks is little affected by the presence of 
expansion and/or temperature evolution. This implies that 
the system's response to the sudden quench is set by an 
internal dynamics scale that is faster than that of the 
expansion and accompanying temperature fall-off rates  
considered here. 
After the structure function is past its peak, the 
system is isotropized, and, after a relatively short time,  
any memory of the initial fast, spinodal-like, long range response 
to the violent quench across the deconfinemnt transition boundary 
disappears. The subsequent evolution is that of (quasi)equilibrium 
evolution of the system as it expands and cools towards its return to 
the confinement phase.  
 
\begin{figure}[h]
\includegraphics[width=\columnwidth]{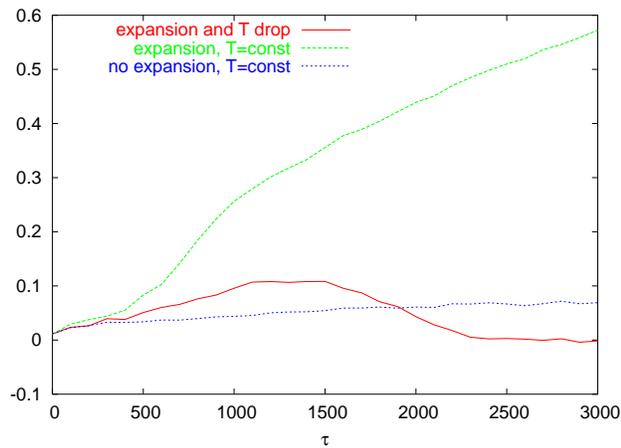}
\caption{The Polyakov loop evolution after 
a quench in decreasing temperature on $32^3\times8$ lattice (same 
initial and final temperature quench as in Fig. 1). \label{fig:dpl1}}
\end{figure}

To elucidate this further we plot the time profile of the Polyakov loop 
average in Fig. \ref{fig:dpl1} for the expanding and non-expanding systems 
as well as with and without drop in temperature. We now have to use 
a bigger, $32^3\times 8$ lattice in order to follow the evolution 
over a longer time interval. 
First note that pure  
expansion drives the Polyakov loop towards larger values, i.e.  
more pronounced deconfined 
behavior. To get some insight into this behavior observe that, as it 
is evident from (\ref{action}),  
spatial expansion enhances the timelike part of the action 
(2nd term in the square brackets on the r.h.s.) while 
suppressing the spacelike parts \cite{F3}.
This tends to increase the 
expectation of timelike Polyakov lines \cite{Fsan}.

On the other hand, decreasing temperature counteracts the expansion effect.   
This is clearly seen in Fig. \ref{fig:dpl1}. Eventually, 
under the combined effects of expansion and temperature fall-off, 
the Polyakov loop expectations drops to zero signaling the return of the 
system to the confinement phase. These qualitative features of 
Fig. \ref{fig:dpl1} are rather generic, being stable under 
changes in the expansion and temperature fall-off rates (cf. Fig. 
\ref{fig:dpl2} below). 
 
The crucial feature characterizing 
the system's overall evolution following the rapid quench into 
the deconfinement region is clearly revealed 
by examining Fig. \ref{fig:dpl1} in conjunction with the 
plots of the structure function in Fig. \ref{fig:tdrop}. It is the fact 
that there are two scales involved in this evolution. The first is the 
scale set by the location of the peak of the structure function; 
the second is  the scale set by the interval to return to the 
confinement phase. Furthermore, {\it there is wide separation between 
these two scales.} As noted above, the location of the peak ($\sim 300$ 
in Fig. \ref{fig:tdrop}) is very little affected by the conditions of 
expansion and temperature  fall-off. 
It reflects strongly coupled dynamics at short time scales 
driving the exponential growth of very long range modes \cite{F4}
leading to `isotropization', by which we mean nothing more specific than that 
the full development of these modes appear 
to completely wipe out any remnants of the quenching  event.  
Rapid equilibration follows within an 
interval of a few hundreds sweeps ($\sim 300-400$ in fig. \ref{fig:tdrop}). 
The system then continues to evolve in quasi-equilibrium over a much 
longer period (typically of thousands of sweeps as in Fig. \ref{fig:dpl1} 
and Fig. \ref{fig:dpl2} below) expanding 
and cooling till it returns to the confinement phase. 
This separation of time scales is a very general feature over a wide range 
of parameters.

Having reached this qualitative physical picture, which accords well 
with that experimentally observed, it is interesting to explore 
whether it is possible to make an 
estimate of this separation in physical units,  and establish 
some correspondence with heavy-ion collision phenomenology. 
This we do in the following subsection. 

\subsection{Isotropization-thermalization and chemical freeze-out 
times}
The expansion and temperature variation parameters 
($\tau_0^\prime$, $\tau_0$, $T_{\rm final}$)  
determine the precise time interval before returning to the 
confinement phase. Note that our requirement that the evolution remain 
inside the scaling window limits the time of observation.  
To deal with this and follow the evolution for substantial time we 
now work on a larger, $32^3$x$8$ lattice throughout, and attempt 
to set the parameters so as 
to reflect conditions in heavy ion collision at RHIC. There is strong 
evidence that the matter in the firetunnel 
reaches temperatures above $2T_c$ \cite{Kolb:2003dz}. For our numerical 
simulation we quench to $T_{\rm final} \sim 3T_c$. (Note that $T_c$ here means
critical temperature of pure gluodynamics).
The system is equilibrated in the confinement phase 
at the same temperature as before $\approx0.8T_c$.
For $N_\tau=8$ and  $\xi=1$ we recalculate the values of corresponding betas 
using the two-loop order formula connecting lattice spacing and 
the coupling, and reweigh it with non-perturbative correction 
factor above $T_c$ \cite{Boyd:1996bx}. Such a correction is 
needed here since the bare gauge coupling is typically of order one 
for our range of betas. The value of $\beta_c$ is known from the lattice 
Polyakov loop susceptibility study \cite{Boyd:1996bx}.
We get
\begin{eqnarray}
	\beta_{\rm initial}&=5.90 &\quad 0.76T_c\nonumber\\
	\beta_c&=6.0625 &\quad T_c\\
	\beta_{\rm final}&=6.85 &\quad 2.95T_c\nonumber\;,
\end{eqnarray}
where `initial' and `final' refer to before and after the quench.

\begin{figure}[ht]
\includegraphics[width=\columnwidth]{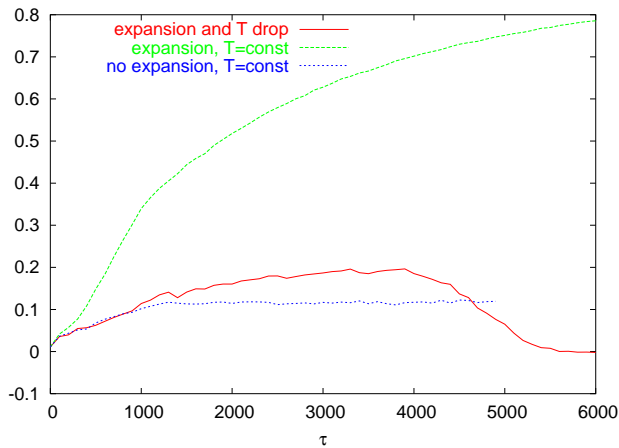}
\caption{The Polyakov loop evolution after a quench to $2.95T_c$, 
$\tau_0=1000$ \label{fig:dpl2}}
\end{figure}
At high enough temperatures the system is
close to an ideal gas of quasiparticles.  
Lattice data suggest that the equation of state 
does not actually deviate much from that of 
the ideal gas for all temperatures down to $T_c$, and is 
conveniently modeled as such in hydrodynamic descriptions. 
Hence, the energy and entropy densities  scale as $e\sim T^4$ and 
$s\sim T^3$, respectively. Assuming adiabatic expansion the entropy 
per unit of rapidity is conserved. In real fire-tunnel evolution 
the initial expansion is one dimensional \cite{Bjorken:1983}, 
switching to three dimensional expansion at later times. This corresponds to 
$T\sim\tau^{-1/3}$ at the early and mid time and 
$T\sim\tau^{-1}$ at later time. The $T\sim\tau^{-1/3}$ behavior is 
in fact seen in the hydrodynamic evolution almost down to 
$T_c$ \cite{Heinz:2004pj,Kolb:2003gq,Kolb:2003dz}, and is the 
only one considered in the following. 
We are thus led to modeling of 
the spatial expansion and temperature drop on the lattice in 
terms of the two parameters $\tau_0$ and $\tau_o^\prime$ as: 
\begin{eqnarray}
a_s&=&a_{0}(1+\frac{\tau}{\tau_{0}}),\\
a_\tau&=&a_{0}(1+\frac{\tau}{\tau_{0}^\prime})^{1/3} \nonumber \\
&=&a_{0}(1+y\frac{\tau}{\tau_{0}})^{1/3}, 
\end{eqnarray}
where $y=\tau_0/\tau^\prime_0$ is the ratio of the `speeds'. The 
evolution of the anisotropy then is 
\begin{equation}
	\xi(\tau) = \frac{1 +{\tau\over \tau_0}} 
           {(1+ y\,{\tau\over \tau_0})^{1/3} } 
\end{equation}

As the chemical freeze-out temperature we choose $T_{fo}=T_c$ 
\cite{Kolb:2003dz,Heinz:2004pj,Kolb:2003gq}.
Before this freeze-out the plasma undergoes an $x$-fold expansion
\begin{equation}
	x=\frac{a_s}{a_0}=1+\frac{\tau_{fo}}{\tau_0}
	\label{eq:x_def}
\end{equation}
On the other hand
\begin{equation}
	\frac{T_{fo}}{T_{final}}=
	\frac1{(1+y\frac{\tau_{fo}}{\tau_0})^{1/3}}
\end{equation}
From these two we get
\begin{equation}
	y=\frac{(T_{final}/T_{fo})^3-1}{x-1}
\end{equation}

Hydrodynamical model phenomenology yields all required parameters 
\cite{Kolb:2003gq}. Thus we have $x\sim9$ before the conversion to the
confined phase is completed, and the freeze-out is observed. 
This corresponds to $y\sim3.25$.

In Fig. \ref{fig:dpl2} we plot the time profile of the Polyakov loop 
average in expanding and non-expanding systems with and 
without drop in temperature with these values of $x$, $y$, and $\tau_0=1000$.  
One again observes the same features discussed above in connection 
with Fig. \ref{fig:dpl1}. 

\begin{figure}[h]
\includegraphics[width=\columnwidth]{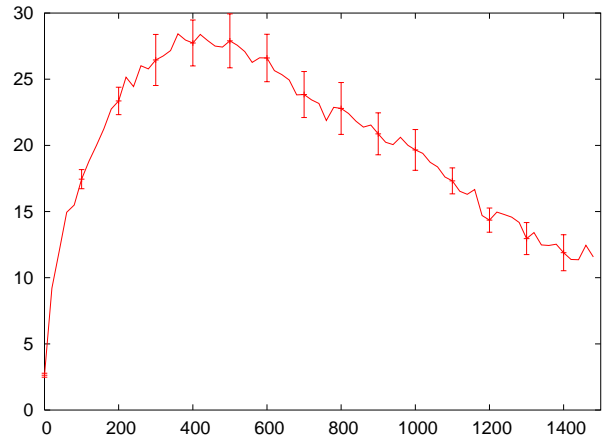}
\caption{The first mode of structure function for  
non-expanding and constant $T$ system after a quench.\label{fig:sf1_noet}}
\end{figure} 
In Fig. \ref{fig:sf1_noet} we  show the lowest mode structure function 
in the case of no expansion and temperature variation. The 
expanding-variable temperature case is 
very similar. We see that at $\tau=400$ sweeps
the system has reached the peak.
This is the point of isotropisation of the system when the long range 
fluctuations reach through the system.
It is again to be contrasted to the much longer times needed to 
return to confinement (vanishing Polyakov loop expectation). 
\begin{figure}[ht]
\includegraphics[width=\columnwidth]{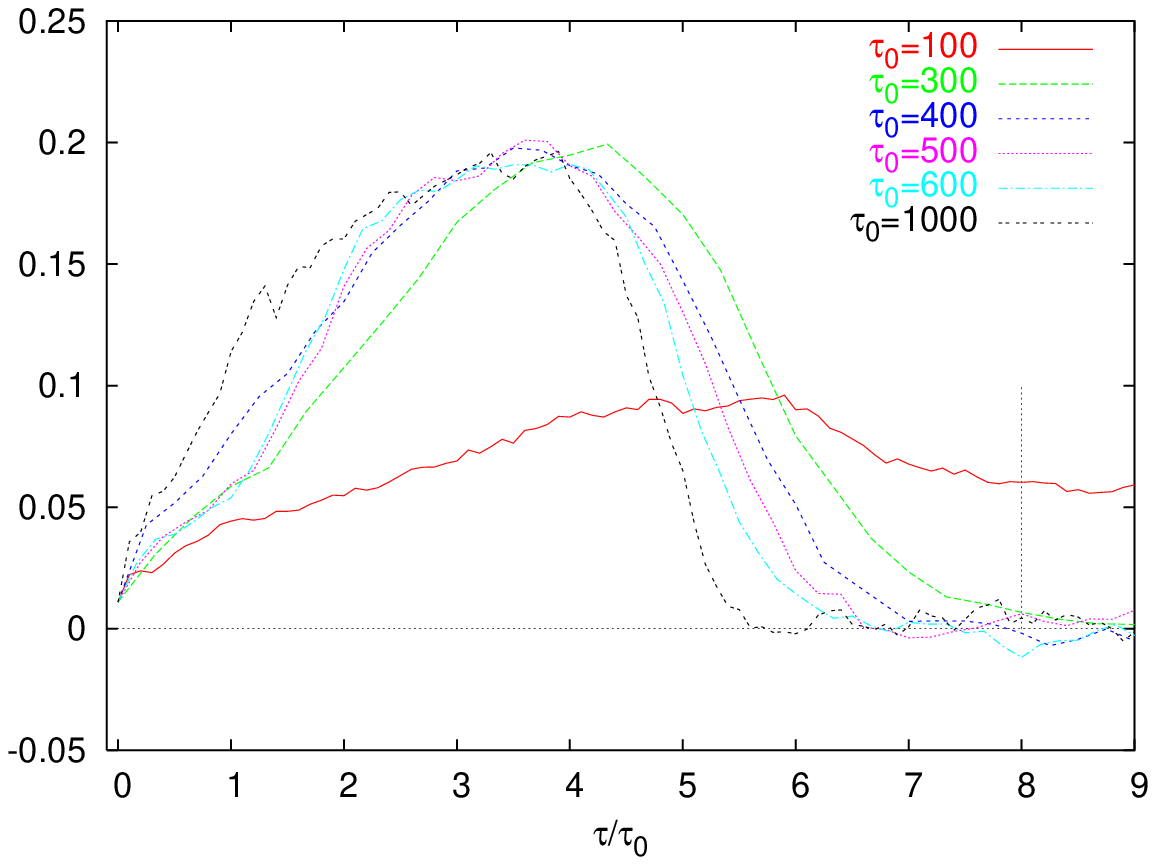}
\includegraphics[width=\columnwidth]{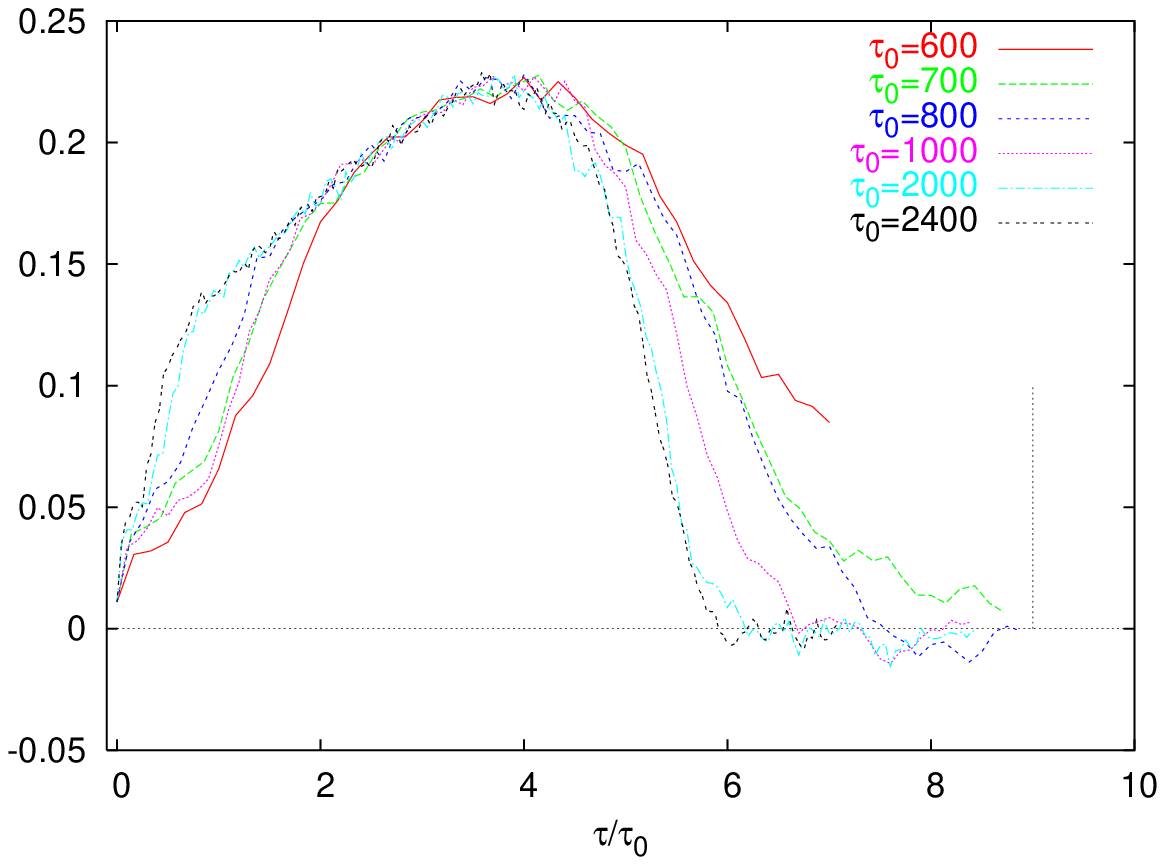}
\caption{The Polyakov loop evolution after a quench to $2.95T_c$ 
for various values of $\tau_0$ and  
$x=9$ (top panel), $x=10$ (bottom panel).\label{fig:dpl_d}} 
\end{figure}

To explore this difference in detail, in  Fig. \ref{fig:dpl_d} we
present the Polyakov loop average evolution at different expansion rates. 
The figure on the top panel uses again $x=9$, so 
the chemical freeze-out point on this plot is at $\tau/\tau_0=(x-1)
=8$. We notice that in the range of $\tau_0\sim 400-600$ 
the Polyakov loop expectation value gets close to zero just before 
this point. Actually, this should be considered as a lower bound 
on the range of $\tau_0$'s since this is a first order transition  
(or a  rapid crossover in the presence of fermions). Thus, there is a 
latent heat period during which the system lingers, with Polyakov loop 
expectations close to zero, before it is fully converted to the 
confined phase. 
The values $\tau_0=300$ and less (higher `speeds') do not show this 
behavior and therefore do not lead to a consistent picture. 
Taking then the range of $\tau_0 \gtrsim 400-600$ gives 
an interval to freeze-out of $\gtrsim 3200-4800$ sweeps, which in turn 
corresponds, from phenomenology, to $\sim 9 fm/c$. 
This allows an upper bound estimate of the time around the peak of the 
structure function in physical units: $\lesssim 0.75-1.1\ fm/c$. 
 
In the plot on the bottom panel in Fig. \ref{fig:dpl_d} we use the 
same quench to $2.95 T_c$ but now take $x=10$, which gives $y=2.89$. This 
corresponds to a somewhat longer lifetime for the deconfined plasma. 
One sees that a value of $\tau_0 \gtrsim 700 - 800$ gives a good fit 
to complete conversion to the confined phase by the time  
the  freeze-out point is reached. This in turn gives 
$\lesssim 0.56 - 0.63\ fm/c$ as an upper bound estimate for the time around 
the peak of the structure function. 
There is of course an inherent  uncertainty  here as to what to take for 
the appropriate phenomenological value for the lifetime in physical units 
since we do not include fermions in out simulations; and  
fermions become important at the late stage of conversion back to confinement. 
In Fig. \ref{fig:dpl_d} 
we follow the curves as far as possible before running out of the 
scaling regime as explained in Section 3.1 above. Longer lifetimes 
result into even shorter isotropization times. To consider somewhat 
larger $x$ values, however, we need larger lattices than those employed in 
this study. But the main message extracted from the present decimations 
should be clear. The robust separation between the 
`fast' dynamics scale of the exponential growth of the 
spinodal-like response to the sudden quench  and that set by 
return to confinement is such that 
isotropization times well inside $1 \ fm/c$ result naturally 
for any reasonable choice of parameters.

\section{Conclusions}
In this paper we studied the response of the pure $SU(3)$ gauge theory 
to a rapid quench from its confined to its deconfined phase. In a 
series of simulations we followed the subsequent evolution of the system 
under varying conditions of temperature fall-off and/or spatial expansion.  
These conditions were chosen so as to reflect the type of 
variations presumed to hold in heavy ion collisions. 
Our main finding is that there are two distinct scales characterizing 
this evolution. There is one scale set by the development of 
very long range modes continuously from zero to their maximum 
amplitude over a short time interval. These modes, manifested in the 
response of the structure functions to the quench, drive 
isotropization and return to thermalization shortly afterwards. 
The scale set by this `fast' dynamics is little affected by 
conditions of expansion and temperature variations.  The second scale 
is set by the time interval to return to the confinement 
phase under quasi-equilibrium evolution, and is affected by the details  
of expansion and temperature fall-off. There is a robust wide 
separation between the  two scales, which,  
translated to physical units under reasonable assumptions 
about the lifetime of the plasma, gives estimates of 
the `fast' dynamics in the range of $0.5 - 1 \ fm/c$. 

There is a number of directions in which this study can be extended. 
The formalism used above can be extended to treat also anisotropy among the 
different space directions, thus allowing a more `realistic' treatment of 
spatial expansion.  Whether this makes much of difference, however, 
remains to be seen (cf. footnote \cite{Fsan}). 
Another interesting question is that concerning the effect of having a 
really finite physical system. In our simulations we, as usual, 
employed periodic boundary conditions. But one may explore different 
ones that would mimic a finite rather than an infinite system.  
The other obvious extension is the inclusion of fermions. The 
deconfinement transition is driven by gluonic dynamics. 
Fermions, however, are expected to contribute more substantially 
at the late stage of return to confinement and hadronization. 
Thus our study applies strictly to the gluonic plasma. 
Nonetheless, the qualitative picture of the separation 
of scales found above should not be 
affected in an essential way by the inclusions of fermions, though 
quantitative details related to the exact lifetime of the plasma, etc  
certainly will.

\section*{Acknowledgment}
We thank Academical Technology Services at UCLA for computer support.
Our code is build on top of the SciDac qdp++ library, which
also forms the foundation for chroma \cite{Edwards:2004sx}. 
We also like to thank B. Berg for discussions. This work is 
partially supported by NSF-PHY-0309362.


\begin{thebibliography}{99}
\bibitem{Heinz:2004pj}
  U.~W.~Heinz,
  %``Thermalization at RHIC,''
  AIP Conf.\ Proc.\  {\bf 739}, 163 (2005)
  [arXiv:nucl-th/0407067].
  %%CITATION = NUCL-TH 0407067;%%
\bibitem{Kolb:2003gq}
  P.~F.~Kolb,
  %``Expansion rates at RHIC,''
  Heavy Ion Phys.\  {\bf 21}, 243 (2004)
  [arXiv:nucl-th/0304036].
  %%CITATION = NUCL-TH 0304036;%%
\bibitem{Kolb:2003dz}
  P.~F.~Kolb and U.~W.~Heinz,
  %``Hydrodynamic description of ultrarelativistic heavy-ion collisions,''
  arXiv:nucl-th/0305084.
  %%CITATION = NUCL-TH 0305084;%%
\bibitem{MG} D. Molnar and M. Gyulassy, Nucl.\ Phys.\ A {\bf 697}, 495 (2002). 
\bibitem{BjV} J. Bjoraker and R. Venugopalan, Phys. Rev. C {\bf 63}, 
  024609 (2001).
\bibitem{BMSS} R. Baier, A.H. Mueller, D. Schiff, D.T. Son, 
  Phys.\ Lett.\ B {\bf 539}, 46 (2002); ibid. {\bf 502}, 51 (2001). 
\bibitem{Kar} D. Kharzeev, arXiv:hep-ph/0408091. 
\bibitem{IV} E. Iancu and R. Venugopalan, arXiv:hep-ph/0303204. 
\bibitem{Kov} Y.V. Kovchegov, arXiv:hep-ph/0503038; arXiv:hep-ph/0507134.  
\bibitem{ALM} P. Arnold, J. Lenaghan and G.D. Moore, JHEP {\bf 0308}, 002 
  (2003); P. Arnold and J. Lenaghan, Phys. Rev. D {\bf 70}, 114007 (2004). 
\bibitem{RS1} P.~Romatschke and M.~Strickland, 
Phys.\ Rev.\ D {\bf 68}, 036004 (2003) [arXiv:hep-ph/0304092]. 
\bibitem{M} S. Mrowczynski, Phys.\ Lett.\ B {\bf 393}, 26 (1997); 
  {\bf 314}, 118 (1993); Phys. Rev. C {\bf 49}, 2191 (1994). 
\bibitem{H} U.W. Heinz, Nucl.\ Phys.\ A {\bf 418}, 603C (1984). 
\bibitem{ALMY} P. Arnold, J. Lenaghan, G.D. Moore and L.G. Yaffe, 
  Phys.\ Rev.\ Lett.\  {\bf 94}, 072302 (2005). 
\bibitem{Rebh1} A.~Rebhan, P.~Romatschke and M.~Strickland,
 %``Hard-loop dynamics of non-Abelian plasma instabilities,''
 Phys.\ Rev.\ Lett.\  {\bf 94}, 102303 (2005) [arXiv:hep-ph/0412016].
 \bibitem{RS2} P.~Romatschke and M.~Strickland,
 %``Collective modes of an anisotropic quark-gluon plasma. II,''
 Phys.\ Rev.\ D {\bf 70}, 116006 (2004) [arXiv:hep-ph/0406188]; 
 S.~Mrowczynski, A.~Rebhan and M.~Strickland,
 %``Hard-loop effective action for anisotropic plasmas,''
 Phys.\ Rev.\ D {\bf 70}, 025004 (2004) [arXiv:hep-ph/0403256].  
\bibitem{DN} A. Dumitru and Y. Nara, arXiv:hep-ph/0503121.
\bibitem{MM} C. Manuel and S. Mrowczynski, arXiv:hep-ph/0504156.
\bibitem{AMY} P. Arnold, G.D. Moore and L.G. Yaffe, arXiv:hep-ph/0505212.
\bibitem{Rebh2} A.~Rebhan, P.~Romatschke and M.~Strickland, 
arXiv:hep-ph/0505261. 
\bibitem{gunton:1985}
 J.~D.~Gunton and M.~Droz,
 "Introduction to the Theory of Metastable and Unstable States", 
  Springer-Verlag (1985). 
\bibitem{Chaikin:1997}
     P.~M.~Chaikin and T.~C.~Lubensky",
     "Principles of condensed matter physics", Cambridge University Press 
     (1997). 
\bibitem{Pisarski:2000eq}
  R.~D.~Pisarski,
  %``Quark-gluon plasma as a condensate of Z(3) Wilson lines,''
  Phys.\ Rev.\ D {\bf 62}, 111501 (2000)
  [arXiv:hep-ph/0006205].
  %%CITATION = HEP-PH 0006205;%%
\bibitem{W} N. Weiss, Phys.\ Rev.\ D {\bf 24}, 475 (1981); ibid, 
  {\bf 25}, 2667 (1982). 
\bibitem{HKW} T. Heinzl, T. Kaestner and A. Wipf, arXiv:hep-lat/0502013.
\bibitem{gunton:1983}
 J.~D.~Gunton, M.~San\ Miguel and P.~S.~Sahni,
%      editor    = "Domb, C. and Lebowitz, J.L.",
%      %"The Dynamics of First-order Phase Transitions",
      in ``Phase Transitions and Critical Phenomena, v. 8'',  
      C.~Domb and J.~L.~Lebowitz (editors), Academic Press (1983). 
%      address   = "London",
%      year      = "1983"
\bibitem{Miller:2000pd}
  T.~R.~Miller and M.~C.~Ogilvie,
  %``Spinodal decomposition in high temperature gauge theories,''
  Phys.\ Lett.\ B {\bf 488}, 313 (2000)
  [arXiv:hep-lat/0004004].
  %%CITATION = HEP-LAT 0004004;%%
\bibitem{Miller:2000mr}
  T.~R.~Miller and M.~C.~Ogilvie,
  %``Spinodal decomposition in finite temperature SU(2) and SU(3),''
  Nucl.\ Phys.\ Proc.\ Suppl.\  {\bf 94}, 419 (2001)
  [arXiv:hep-lat/0010055].
  %%CITATION = HEP-LAT 0010055;%%
\bibitem{Miller:2001ym}
  T.~R.~Miller and M.~C.~Ogilvie,
  %``Nonequilibrium aspects of quantum field theory,''
  Nucl.\ Phys.\ Proc.\ Suppl.\  {\bf 106}, 537 (2002)
  [arXiv:hep-lat/0110109].
  %%CITATION = HEP-LAT 0110109;%%
\bibitem{Bazavov:2004wc}
  A.~Bazavov, B.~A.~Berg and A.~Velytsky,
  %``Model A dynamics and the deconfining phase transition for pure lattice
  %gauge theory,''
  arXiv:hep-lat/0410019.
  %%CITATION = HEP-LAT 0410019;%%
\bibitem{Berg:2003mn}
  B.~A.~Berg, U.~M.~Heller, H.~Meyer-Ortmanns and A.~Velytsky,
  %``Spinodal decomposition and the deconfining phase transition,''
  Nucl.\ Phys.\ Proc.\ Suppl.\  {\bf 129}, 587 (2004)
  [arXiv:hep-lat/0308032].
  %%CITATION = HEP-LAT 0308032;%%
\bibitem{Berg:2003hc}
  B.~A.~Berg, U.~M.~Heller, H.~Meyer-Ortmanns and A.~Velytsky,
  %``Dynamics of phase transitions by hysteresis methods. I,''
  Phys.\ Rev.\ D {\bf 69}, 034501 (2004)
  [arXiv:hep-lat/0309130].
  %%CITATION = HEP-LAT 0309130;%%
\bibitem{Berg:2004qb}
  B.~A.~Berg, H.~Meyer-Ortmanns and A.~Velytsky,
  %``Dynamics of phase transitions: The 3D 3-state Potts model,''
  Phys.\ Rev.\ D {\bf 70}, 054505 (2004)
  [arXiv:hep-lat/0405011].
  %%CITATION = HEP-LAT 0405011;%%
\bibitem{Velytsky:2002fn}
  A.~Velytsky, B.~A.~Berg and U.~M.~Heller,
  %``Dynamics of the 2d Potts model phase transition,''
  Nucl.\ Phys.\ Proc.\ Suppl.\  {\bf 119}, 861 (2003)
  [arXiv:hep-lat/0208035].
  %%CITATION = HEP-LAT 0208035;%%
\bibitem{Cahn:1958}
     J.~W.~Cahn and J.~E.~Hilliard,
     %``Free Energy of a Nonuniform System. I. Interfacial Free Energy'',
     J.\ Chem.\ Phys.\ {\bf 28}, 258 (1958).
\bibitem{Velytsky:2004}
  A.~Velytsky, Ph.D. thesis,
%     title     = "A model study of the deconfining phase transition",
   UMI-31-37501
\bibitem{F1}
An alternative dynamical model based on an   
extension of (\ref{Polyak-pot}) that includes a chiral field,  
and involving second-order time derivatives was   
introduced in \cite{Scavenius:2001pa}. It was used to study the transition  
back into the confinement phase at the end of the QGP evolution,  
in contrast to our first-order model used here to study the 
initial response right after the quench into the deconfinement phase.
\bibitem{Scavenius:2001pa}
  O.~Scavenius, A.~Dumitru and A.~D.~Jackson,
  %``Explosive decomposition in ultrarelativistic heavy ion collision,''
  Phys.\ Rev.\ Lett.\  {\bf 87}, 182302 (2001)
  [arXiv:hep-ph/0103219].
  %%CITATION = HEP-PH 0103219;%%
\bibitem{Bjorken:1983}
  J.~D.~Bjorken,
  %``Highly Relativistic Nucleus-Nucleus Collisions: The Central Rapidity
  %Region,''
  Phys.\ Rev.\ D {\bf 27}, 140 (1983).
  %%CITATION = PHRVA,D27,140;%%
\bibitem{Karsch:1982ve}
  F.~Karsch,
  %``SU(N) Gauge Theory Couplings On Asymmetric Lattices,''
  Nucl.\ Phys.\ B {\bf 205}, 285 (1982).
  %%CITATION = NUPHA,B205,285;%%
\bibitem{F3}
The effect is 
qualitatively similar to that of having no spatial expansion but, instead, 
raising the temperature.

\bibitem{Fsan}
We parenthetically remark that this  observation leads to the 
expectation that, as far as 
the behavior of the timelike Polyakov line is concerned, 
the difference between 3-dimensional and 1-dimensional spatial expansion 
may be numerically not all that significant. (Clearly, 
this may be not be the case for other physical quantities.) 
This is because, as seen from (\ref{action}), such a difference affects only 
the degree of expansion-induced suppression of interactions in the different  
space directions, but not the expansion-induced enhancement of the timelike 
interactions; and it is the latter that mainly affects the  
Polyakov loop enhancement under expansion seen in  Fig. \ref{fig:dpl1}.  
Whether this expectation concerning changes in the dimensionality 
of the spatial expansion is indeed correct remains, of course, to be 
decided by actual computations not carried out in this paper.
\bibitem{F4}
It is 
worth noting again that all higher momentum modes are 
essentially unaffected in this type of response to the quench, a 
somewhat counter-intuitive behavior, which, however, is well 
known in spinodal-like responses.
\bibitem{Boyd:1996bx}
  G.~Boyd, J.~Engels, F.~Karsch, E.~Laermann, C.~Legeland, M.~Lutgemeier and B.~Petersson,
  %``Thermodynamics of SU(3) Lattice Gauge Theory,''
  Nucl.\ Phys.\ B {\bf 469}, 419 (1996)
  [arXiv:hep-lat/9602007].
  %%CITATION = HEP-LAT 9602007;%%
\bibitem{Edwards:2004sx}
 R.~G.~Edwards and B.~Joo  [SciDAC Collaboration],
 %``The Chroma software system for lattice QCD,''
 arXiv:hep-lat/0409003.
 %%CITATION = HEP-LAT 0409003;%%



\end{thebibliography}
\end{document}